\documentclass[runningheads]{llncs}
\usepackage{graphicx}
\usepackage{url}

\begin{document}

\title{How NFT Collectors Experience Online NFT Communities: A Case Study of Bored Ape}
\titlerunning{How NFT Collectors Experience Online NFT Communities}

\author{Allison Sinnott\inst{1} \and
Kyrie Zhixuan Zhou\inst{2}}
\authorrunning{Sinnott and Zhou}
\institute{
East Meadow High School, East Meadow, New York \\
\email{sial2706@emufsd.org}\\
\and
University of Illinois at Urbana-Champaign, Champaign, Illinois\\
\email{zz78@illinois.edu}}

\maketitle           

\begin{abstract}
Non-fungible tokens (NFTs) are unique cryptographic assets representing the ownership of digital media. NFTs have soared in popularity and trading prices.
However, there exists a large gap in the literature regarding NFTs, especially regarding the stakeholders and online communities that have formed around NFT projects. Bored Ape Yacht Club (BAYC) is one of the most influential NFT projects. Through an observational study of online BAYC communities across social media platforms and semi-structured interviews with four participants who owned BAYC NFTs, we explored the experiences of NFT collectors within the online NFT community. Positive community experiences, i.e., personal expression and identity, mutual support among BAYC holders, and exclusive access to online and offline events, were expressed.
Encountered challenges included scams and ``cash grab'' NFT projects as well as trolling.
The results of this study point towards the welcoming, positive nature of the NFT community, which is a possible causation factor of the initial rise in popularity of NFTs. Demotivators, on the other hand, countered the established trustworthiness of NFT technology among its consumers.
\end{abstract}

\section{Introduction}
\label{sec:introduction}

\begin{figure}
{\includegraphics[height=5cm]{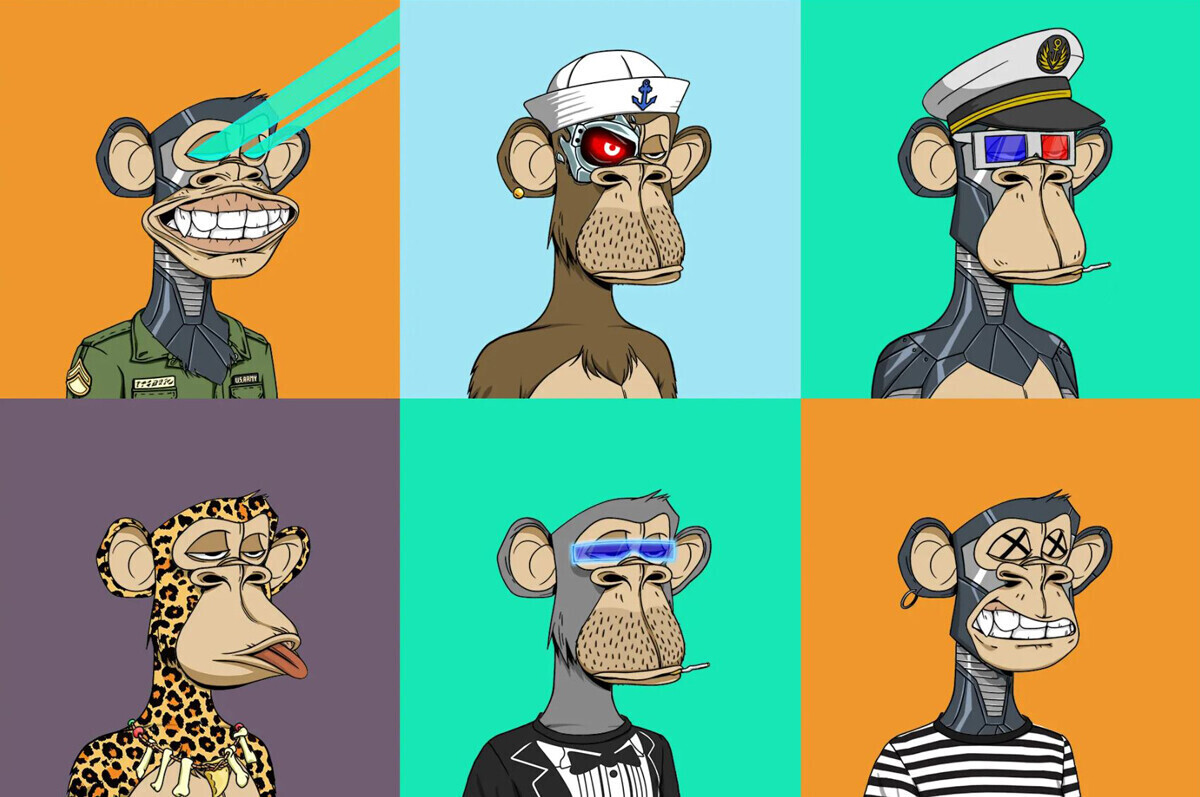}}
\centering
\caption{Various Bored Ape NFTs. Source images courtesy Bored Ape Yacht Club.}
\label{fig:bayc}
\end{figure}

A non-fungible token (NFT) is a unique cryptographic asset representing the ownership of digital content, such as an image, an audio file, a video, or any other digital asset. These tokens replicate the properties of physical items, such as scarcity and uniqueness \cite{Regner:2019}. NFTs are purchased and valued using cryptocurrencies, decentralized virtual currencies that rely on anonymous peer-to-peer networks that maintain a public ledger of transactions~\cite{Gao:2016}.

The story of NFTs, initially called ``monetized graphics,'' began in 2014~\cite{McCoy:2014}.
The original premise of NFTs was to help artists exercise ownership, distribution, and control of their digital artworks \cite{Dash:2021}. At this time, ``Quantum,'' a digital work of art made by Kevin McCoy, became the first NFT ever released and was sold for an amount of cryptocurrency equivalent to four US Dollars \cite{Cascone:2021,Dash:2021}.
It took seven years for NFT trading to climax.
In March 2021, digital artist Mike ``Beeple'' Winkelmann achieved record revenue for the digital artwork, ``Everydays: The First 5000 Days,'' which was sold for \$69.3 million \cite{Reyburn:2021}.
After this sale, Winkelmann stated, ``I believe we are witnessing the beginning of the next chapter in art history, digital art'' \cite{Weiner:2021}.
Other NFT creators agree with Winkelmann, calling the explosion of NFTs the ``New Renaissance'' \cite{Sharma:2022}.

Despite the explosion in popularity, price, and expected impact of NFTs, little is known about the stakeholders involved in the NFT ecosystem and their perceptions of its development. Prior research regarding NFTs examined the underlying blockchain technology or NFTs' impact on the art world \cite{Franceschet:2021,Regner:2019}.
Other literature revolved around the cryptocurrencies NFTs were valued in and exchanged for~\cite{Khairuddin:2016}.
Sharma et al. investigated NFT creators' perceptions \cite{Sharma:2022}, but no extant literature has touched on the experiences of NFT collectors.

NFT collectors band together in social media communities based on the type of NFTs they collect~\cite{Casale-Brunet:2022}.
Such groups constitute optimal environments to explore stakeholders' perceptions and the impact communities have on the explosive growth of NFTs.
Our study aims to understand the lived experiences of NFT collectors in online communities.
Understanding the formation and dynamics of NFT communities creates grounds for further research into the causes of the NFT explosion.
We use Bored Ape Yacht Club (BAYC), one of the most influential NFT projects \cite{Casale-Brunet:2022}, for a case study. Our research question (RQ) is: \textit{How do Bored Ape Yacht Club (BAYC) collectors perceive the communities that have formed around the NFT phenomenon?}

Through interviews with 4 BAYC collectors, we found they enjoyed multiple benefits brought by BAYC NFTs and the surrounding online community, such as expressing their personality and identity, being granted exclusive access to social opportunities and online/offline events, enjoying the welcoming and supportive community atmosphere, etc. However, negative experiences such as scam NFTs and ``trolling'' (being the subject of ridicule for displaying NFT artwork) were also commonly expressed.
Our observation of online communities confirmed some of the interview results -- people formed a cohesive online community for discussing NFTs, blockchain, and daily lives; scams were prevalent but people acted collectively against them by providing warnings for community members; community usage of IP rights was common.

The contribution of this work is thus two-fold.
Firstly, we unveil BAYC collectors' experiences in the community with the first observational study.
Secondly, we connect our findings to the rises and falls of the NFT market, laying a foundation for more research into the causation factors of the evolution of the NFT space.

\section{Related Work}
\label{sec:related-work}

\subsection{The Underlying Technology of NFTs}
\label{subsec:the-underlying-technology-of-nfts}
Blockchains, the infrastructure for NFT creation and transaction, are anonymous peer-to-peer networks that maintain public ledgers and cryptographic proofs of transactions \cite{zohar:2005}.
Transactions are irreversible once completed.
Blockchain first gained popularity as the new technology behind the cryptocurrency of Bitcoin,
which was proposed as a means for online payments to be sent from one party to another without the need for a middleman or the possibility of state intervention \cite{Nakamoto:2008}. Blockchain, a technology that decentralizes trust in a previously centralized domain, enticed many. However, besides legitimate uses, cryptocurrencies have a history of being instrumental to criminal activity, such as being used for payment in dark web transactions~\cite{Braaten:2019}.

Though blockchain was originally intended to handle monetary transactions, NFTs are an example of blockchain technology applied to a setting beyond cryptocurrency \cite{Conti:2018}.
Blockchain is a particularly suitable underlying technology for NFTs as it has low information privacy requirements due to the public nature of ownership verifiability~\cite{Platt:2021a}.
The Ethereum blockchain, which a lot of NFT projects are built upon, is a publicly accessible, permissionless blockchain. As a proof-of-stake blockchain, Ethereum has significantly lower electricity requirements than its proof-of-work counterparts which have been commonly criticized for their extreme energy demands~\cite{Platt:2021b,dragnoiu2023more}.
The accessibility of the Ethereum blockchain makes it an ideal infrastructure for NFT creators to create and deploy their programs  \cite{Buterin:2014,Wood:2014}, which are also known as smart contracts \cite{sharma2023mixed}. Smart contracts intend to execute actions according to the terms of a contract/agreement \cite{Regner:2019}.
Smart contracts have become the foundational building blocks for NFTs and created confidence in the security of this technology. Blockchains dedicated to NFTs, such as Flow, have also been created \cite{Flow}. With secure blockchain networks created for the widespread public implementation of smart contracts, the stage was set for NFT projects, such as BAYC, to flourish.

\subsection{NFT Marketplaces}
\label{subsec:nft-marketplaces}

NFT marketplaces, such as Opensea\footnote{\url{https://opensea.io/}}, Rarible\footnote{\url{https://rarible.com/}}, and Foundation\footnote{\url{https://foundation.app/}}, can be best compared to art exhibitions or auction houses. Both have two primary stakeholders: the artists and the buyers. In both the NFT marketplace and traditional auction houses, unique, non-fungible works are sold. The value of both NFTs and physical art is based on their rarity and quality as well as the reputation of the creator. The main contrast between the two marketplaces is the transparency of ownership. In traditional art markets, forgery and fraud issues can arise \cite{Franceschet:2021}.
In contrast, ownership details of NFTs are recorded and preserved publicly on the blockchain, allowing anyone to verify the authenticity of ownership.
This feature improves the trustworthiness of NFTs and contributes to the motivation of collectors to buy them.

NFT creators and buyers are the two most prominent stakeholders in the NFT ecosystem. A creator would create an NFT by signing a transaction which is sent to the underlying smart contract. The NFT is then associated with an identifier in the blockchain, which is the evidence of ownership. Ownership details are preserved on the blockchain and transferred to the buyer/collector \cite{Sharma:2022,Hong:2021}.

\subsection{The Bored Ape Yacht Club Community}
\label{subsec:the-bored-ape-yacht-club-community}

The creation of NFT projects is accompanied by the formation of online communities. Online communities form around similar interests \cite{iriberri2009life}, which is also the case in the NFT space. Online communities for NFT collectors exist on social media and forums such as Discord channels for popular NFT projects, the NFT subreddit on Reddit, and dedicated Twitter accounts for NFT projects \cite{Chayka:2021,Collins:2022}.

BAYC NFTs were considered by previous literature as a leading force in the NFT marketplace and on social media platforms, having the highest index of posts on Twitter out of all NFT communities \cite{Casale-Brunet:2022,Chayka:2021}. This aspect was the justification for using the BAYC community for our study. The BAYC is an NFT project launched by Yuga Labs, consisting of ten thousand unique variations of cartoon images of multicolored and accessorized apes. Owning one of these NFTs entails membership in the BAYC which grants access to in-person events, exclusive social media channels, and ``the Bathroom,'' a collaborative art experiment which is open only to members \cite{BAYC:2021}. Essentially, ownership of a BAYC NFT grants the collector access to a community of like-minded individuals who share a similar interest. The BAYC provides a united space for its members, creating another motivation to buy a BAYC NFT and engage in the community.

This work explores NFT collectors', in particular, BAYC collectors' motivations for engaging in the online NFT community and the effects community engagement has on the growth of the NFT phenomena.

\section{Methodology}
\label{sec:methodology}
Preconceived notions are difficult to form regarding NFTs as there is poor coverage of literature on this topic, especially regarding NFT collectors'
perspectives. Thus, we took a grounded theory approach \cite{Oktay:2012} toward understanding BAYC collectors' perceptions of NFT communities.
Semi-structured interviews were utilized alongside a content analysis of the BAYC communities on Discord, Reddit, and Twitter, to ensure a broad range of data collection.

\subsection{Interview}
\subsubsection{Participants}
\label{subsec:pre-data-collection}
As our research goal was to understand perceptions of BAYC collectors regarding NFT communities, we sought to recruit individuals who owned a BAYC NFT, were over the age of eighteen, and were comfortable sharing their experiences in the online NFT community. An abundance of NFT creators and collectors can be found on online social media such as Twitter and Discord \cite{Casale-Brunet:2022}.
Therefore, our recruitment focused on those platforms. Direct/private messages on Twitter and Discord were sent to those who have publicly claimed to own a BAYC NFT. An announcement was also composed for the BAYC Discord and Twitter communities to announce recruitment for potential participants.
In these messages and announcements, the researchers introduced themselves and a summary of the study's aim. Individuals who have voluntarily filled out the screening survey and met our recruitment criteria\footnote{The ownership of a BAYC NFT was verified via cross reference of the Ethereum blockchain transaction history and the NFT's metadata in the owner's crypto wallet.} received the consent form to electronically sign, and were informed their participation in the interview would take roughly one hour of their time. 

In the end, four participants engaged in this IRB-approved study in April 2023. Table~\ref{table:1} contains specific participant demographic data. The participants chosen represented a skewed sample, as all participants identified as male. However, no mention of gender was present in the data collected.
Recruitment of only male participants may point to a larger gender disparity in the BAYC community and the blockchain space in general \cite{Hasso:2019}.

\begin{table*}[thb]
\caption{Overview of Interviewees.}
\label{table:1}
\centering
\begin{tabular}{lllll}
\textbf{Alias} & \textbf{Age} & \textbf{Gender} & \textbf{Highest Level of Education} & \textbf{Interview Method} \\
P1 & 46 & Male & Graduate & Discord Voice Chat \\
P2 & 29 & Male & Bachelors of Science & Discord Voice Chat \\
P3 & 49 & Male & Some graduate & Twitter Direct Message \\
P4 & 42 & Male & Some graduate & Discord Voice Chat
\end{tabular}
\end{table*}

\subsubsection{Semi-structured Interview}
\label{subsec:data-collection}
Data collection began with the set of participants who consented and met the sample criteria. Data collection occurred through four one-on-one interviews through Discord and Twitter. One interview took place over Twitter direct message (DM) due to a participant's discomfort with the usage of audio recording.

Both the Discord and Twitter interviews began with the lead author introducing themself and the purpose of the study. Participants were then asked to introduce themselves through an alias, age, gender identity, and education level. Questions went in-depth and asked about participants' experiences with NFTs and BAYC, with a focus on the community aspect.
Example questions included ``How did the community aspect influence your choice of purchase?'' ``Have you observed common topics, values, and norms in the BAYC community?'' The interview protocol is located in Appendix~\ref{interview}; however, the order of questions or the questions themselves were not concrete due to the semi-structured nature of the interviews \cite{Newcomer:2015}. After about one hour of discussion, the interviews concluded. The interviews were manually transcribed verbatim for analysis. Once the transcription was complete, data analysis began.

\subsubsection{Data Analysis}
\label{subsec:data-analysis}
We used an adapted version of the grounded theory approach \cite{Walker:2006} for analysis. The word ``adapted'' is used to describe the method because while an open coding approach was used to analyze transcript data, we did not aim to create a theory in the field. Instead, we only intended to uncover overarching themes in the data and their implications.

\begin{figure*}[thb]
    \centering
	\includegraphics[width=0.9\linewidth]{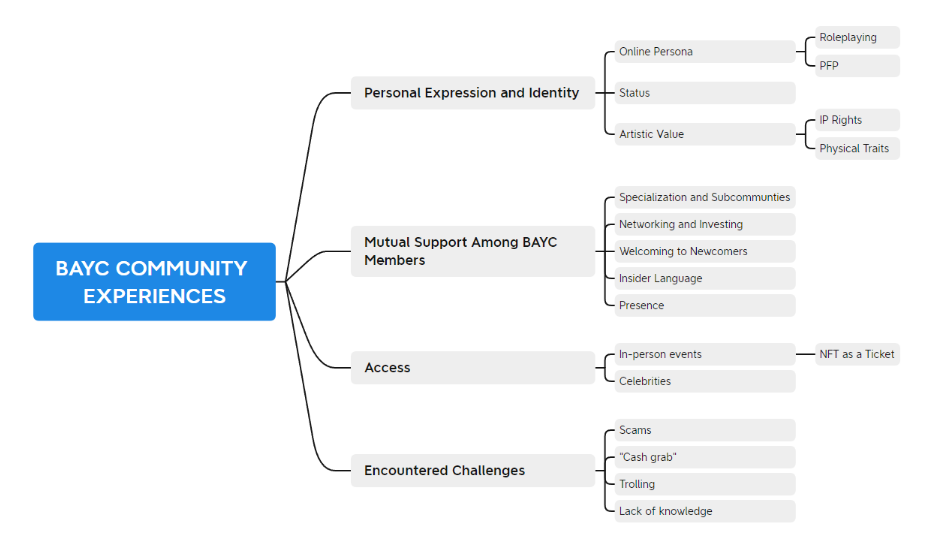}
	\caption{Makeup of themes through codes.}
	\label{mind:mapping}       
\end{figure*}

Coding took place in rounds. The first round of coding was to familiarize the researchers with the data without imposing an interpretation. Subsequent rounds first formed codes, and then found specific patterns, themes, and connections across the data \cite{Lochmiller:2021}. We further used mind mapping to organize emerging themes into a hierarchical structure. Figure~\ref{mind:mapping} is a demonstration of how codes created specific themes.

We use quotes from participants to illustrate our findings. All quotes were anonymized, with participants' aliases used in the report.

\subsection{Participant Observation}
We conducted a participant observation after the interview study for two purposes. First, it was challenging to recruit BAYC holders. We sent out 126 recruitment messages but only four participants accepted the interview request. The observational study supports and generalizes some interview results. Second, by immersing ourselves in the BAYC community, we were better able to observe BAYC holders' behaviors, interactions, and practices in the community on a personal level. The observational study thus served as an ideal complement to the interview study. Ours is the first observational study toward understanding NFT communities.

\subsubsection{Data Collection}

The authors participated in and observed discussions of the BAYC communities on Discord, Reddit, and Twitter for a consecutive week in late August 2023. They took note of the behavior (e.g., promoting BAYC), language (e.g., appraisal of others' BAYC NFTs), norms (e.g., collective resistance to scams), and values (e.g., FOMO) within the BAYC communities, and also conducted informal, unstructured, written interviews with individual BAYC holders to gain further insights into their thoughts and lived experiences.

\subsubsection{Data Analysis} 

A thematic analysis \cite{braun2012thematic} was applied to analyze the notes taken during the participant observation. An open coding process was not used for the observation notes as the interview study had already familiarized us with some phenomena and norms in the BAYC community, so a thematic analysis was more feasible and efficient.

\section{Findings from Interviews}
\label{sec:findings}
The narratives collected through interviews outlined the types of experiences that motivated and demotivated BAYC NFT collectors' interaction with the surrounding community. BAYC members were able to attend events or encounter benefits that were only accessible to BAYC members. BAYC collectors used their NFT or place in the community to express their personality and creativity. A supportive social demeanor was found among members of the BAYC community. However, negative experiences that hindered interaction with others in the BAYC community and demotivated members from further engagement were expressed.

\subsection{Personal Expression and Identity}
\label{subsec:personal-expression-and-identity}
Community members expressed themselves with their NFTs. Two participants, P3 and P4, asserted that art was a major part of their community identity.
P3 wrote on this:

\textit{``The art was a MAJOR factor of WHICH BAYC I selected. I wanted something that would be my `forever ape' and would be something I could make solid use of the IP rights granted to me as an owner of the PFP [profile picture].''}

While P2 did not personally relate with the art of his ape, he explained that the appearance of an NFT was usually to display the owner's personality and interests:

\textit{``People will identify more with NFTs that have similar traits with themselves. Some people smoke cigarettes, some people smoke weed. They often will try and find one bored ape with such traits. Or one with the red eyes trait.''}

Status is a large part of identity in the online NFT space. Ownership of an ape became a status symbol to some. P4 stated:

\textit{``Owning an ape put me on a superstar level and I found that people immediately looked up to me or simply assumed I knew my way around the space because I held an ape.''}

A particularly interesting way of using the BAYC for expression was through the use of role-playing.
Some individuals in the community would use their ape as a vehicle for storytelling. For example, P2 explained the presence of a member who made waves in the community -- the owner of the BAYC character ``Jenkins'' was a New York Times bestselling author who created a collaborative project for other BAYC members to create stories for their apes. P2 has not participated in role-playing or used the creative license for his ape, but recalled that ``it was still definitely one of the cooler uses of community-made IPs.''

\subsection{Exclusive Access}
\label{subsec:exclusive-access}
With the ownership of a BAYC NFT, our participants highlighted the opportunity it gave them to expand their social network and encounter people and experiences that were previously inaccessible. P4 was grateful for these benefits:

\textit{``So just like, being out and active in person in the community has given me experiences to meet and talk to people. I normally wouldn't be able to do so.''}

Two participants, P1 and P2, explored their experiences regarding the founders of the BAYC, Yuga Labs. P2 described how owning an Ape allowed him to attend in-person community meetups:

\textit{``I guess one thing that I can highlight is through these meetups I have been able to meet the two main founders, Gordon and Garga, and I've been able to talk to them a little bit, and kind of express my gratitude to the community that they built.''}

P1 compared this relationship with the founders to the relationship between stockholders and their companies. To P1, a stockholder did not have a say in the direction of their company; it was only an investment to them. With his NFT, however, P1 felt a closer connection to the founders and that he had a say in the direction of the NFT project.

When asked about popular media influence on the community, P2 shared an encounter at ApeFest, an annual music festival open only to Bored Ape holders, with celebrities he admired:

\textit{``I guess the main ones would be Eminem and Snoop Dogg, and I was able to see them at ApeFest, which is really cool. [...] It did kind of help cement that, that I made a good choice or that this community was a good community for me.''}

\subsection{Mutual Support Among BAYC Collectors}
\label{subsec:mutual-support-among-bayc-collectors}
The most prevalent theme in the interviews was mutual support among BAYC collectors. All four participants cited experiences under this theme as being significant pull factors in the online NFT community. P3 reminisced on past interactions with fellow members:

\textit{``I'm only remembering it now -- it's been a while -- but you should check out my Mistaken\_President [Twitter] posts from Dec 2021. I really connected with the community and published a regular update on the floor prices, lol. I really liked vibing with everyone and yeah, it had an impact on my wanting to buy a BAYC.''}

Mutual support represented itself in many ways.
P4 was pleased to share the positive online response from the community when he first bought his ape:

\textit{``They were bombarding your Twitter, with all these likes and replies and things like that. The community made you feel, like, really welcome. They really made their presence felt and I just got sucked into that.''}

The welcoming nature of the community was a large part of P4's decision to join the community long-term.

\subsection{Encountered Challenges}
\label{subsec:encountered-challenges}
Participants largely cited the positive, collaborative, and social aspects of the BAYC community. However, two out of the four participants, P1 and P2, stated challenges they encountered when interacting with the BAYC community and the general online NFT space.

Scams and ``cash grab'' NFT projects are frequently spread to BAYC holders on both Twitter and Discord.
A pump-and-dump is a form of price manipulation that involves artificially inflating an asset's price, in this case, an NFT, before selling the cheaply purchased asset at a higher price. Once the assets are ``dumped,'' the price falls and investors lose money \cite{Li:2021}. P1 explained his frustrations with NFT community members who promoted pump-and-dump NFT projects:

\textit{``There are a lot of bad actors that take advantage. There have been quick cash grabs, and it's kind of them asking somebody who's not really knowledgeable in NFT. They could get easily manipulated or easily duped into, you know, buying something that dumps afterwards, and they could end up losing all their money.''}

P1 further elaborated on an unofficial derivative NFT project based on the BAYC name, the Lazy Lion Yacht Club (LLYC), which caused a large rift in the community. The LLYC collection is a combination of both the BAYC and Lazy Lion NFT collections' visual features. He elaborated:

\textit{``I was part of the side that, you know, saw it as a blatant cash grab and they were using the Bored Ape community, sucking in community members to mint that.''} 

While the LLYC and other unofficial derivative NFT collections did not pretend to be the original NFT or intend to ``phish'' potential buyers, they caused strife in the community nonetheless.

P2 cited another challenge, trolling, as an issue on Twitter and Discord \cite{Morse:2021}. To troll is to make a deliberately offensive online post to upset or elicit an angry response from someone.
He said:

\textit{``Twitter is a different story. There is a lot of trolling. It's just because a lot of people are anonymous on Twitter. A lot of people will kind of troll and not really care about how what they say might affect somebody else. [...] Sometimes there would be certain characters in the Discord that would kind of try and give people a hard time.''}

The decentralized nature of NFTs creates difficulty in combating scams and trolling. However, P2 added that these factors were not much of a discouragement to participating in the BAYC community, but only an annoyance the community must look past.

\section{Findings from Participant Observation}

\subsection{Discord}

\subsubsection{Club Notices}
Channels under the Club Notices category set rules and values for the community.

In the \textbf{\#rules} channel, 10 rules were listed back in 2021, e.g., zero tolerance policy for hate speech, scamming, or other toxic behavior, using appropriate channels for different topics, prohibiting merch trading or reselling, no politics, no unsolicited DMs, no NSFW images, being over the age of 18, etc. During the period of our observation, most rules were well enforced by the moderators and vigilantes in the community. For example, hate speech was not seen; there was no observation of politics-related discussion; people used the appropriate channels for different topics. The remaining challenges turned out to be scamming and unsolicited DMs, which were frequently reported by community members, which we will elaborate on later. 

In the \textbf{\#faq} channel, the moderators outlined common questions and answers concerning what BAYC is, how to become a verified member in the Discord server, etc. Only those who have at least one ape in a MetaMask wallet are recognized as ``Verified Bored Apes'' and provided access to the membership-only channels, echoing the interview results, i.e., exclusive access to social opportunities and events is only granted to those who own an ape. 

Security tips are provided in the form of Q\&A to warn new users about scamming. For example, new users are advised to buy and trade NFTs on OpenSea, a legit NFT marketplace, instead of trading with individual accounts,  

\textit{``If you are new to crypto or NFTs, we honesty suggest you do not even try to trade. Just sell your ape for a `Buy It Now' price on OpenSea and then buy the ape you want. We've seen numerous people get scammed out of their apes, and its heartbreaking.''}

Users are further advised not to click suspicious links, download .SCR files, give seed phrase\footnote{A seed phrase is a sequence of random words that stores the data required to access a crypto wallet.} to anyone, sign suspicious MetaMask transactions, etc. In the decentralized crypto world, people, instead of banks, have to take care of their own crypto assets:

\textit{``Honestly, just assume anyone who is DM'ing you is more than likely a scammer, and act accordingly. This is crypto, and we're all our own bank now. Take security seriously.''}

\subsubsection{User Channels}

There are multiple channels for users to discuss different topics. When one indicated that they were new to the community, other users would direct them to the different channels, showing the welcoming, supportive nature of the BAYC community.

In the \textbf{\#general} channel, people picked up random topics and sent random gifs. In \textbf{\#pet-pics}, people shared pictures of their pets and others reacted with emojis to express their favor. These channels irrelevant to NFTs or crypto suggested the formation of a cohesive community where some people were willing to disclose and chat about their real, offline life. Discord was the most raw and personal level of community interaction, with some of the interaction alike to friends.

In the \textbf{\#crypto-talk} channel, people talked about price trends of BAYC, NFTs, and cryptocurrency in general, as well as the ongoing recession in the economy. With the crypto in the bear market, some expressed the frustration of seeing the BAYC price going down:

\textit{``Yeah I know just hard to be bearish at the bottom at this point if you bought at the Top just keep averaging down.''}

In the \textbf{\#report-scams} channel, people shared information about ape-related scams. At the time of our observation, several users identified an airdrop of Apecoin on Twitter as a scam and shared screenshots of the scam tweets. Some users reported these scam tweets on Twitter: \textit{``yep, already reported thanks.''} Some warned other community members about Discord users who sent scam messages through Direct Message (DM). DM is a common way of scamming. The official account which ran the \textbf{\#bayc-tweets} channel under Club Notices and automatically shared tweets posted by Bored Ape Yacht Club was called ``NGBxBen Will NEVER DM'' with the bio ``I will never DM you first or ask for your seed phrase.'' Such discussions of BAYC scams echoed interview results and illustrated how community members collectively combated the scamming through proactive information sharing and warning.

\subsubsection{Bots}

The Bots \cite{kiene2020uses} channels automatically posted sales activities of BAYC as well as other NFTs by Yuga Labs such as Bored Ape Kennel Club featuring cartoon dog NFTs. The sales activities are linked to their transaction pages in OpenSea. The transaction pages contain such information as the NFTs' price and transaction history, number of views and favorites, and so on. As of August 2023, most apes sold ranged from 25 to 28 ETH, with some reaching 33 ETH, with one ETH equating to around 1,600 US Dollars.

\subsection{Twitter}

\subsubsection{Local Community Accounts}

Accounts across Twitter are dedicated to BAYC members who are looking for a more local, offline community. Twitter accounts have been established for cities across the United States such as @BigAppleApes, based in New York City, @MagicCityApes, based in Miami, and @BoredLApes, based in Los Angeles. However, these local communities are not limited to the United States, where Yuga Labs is based, but span worldwide. International examples include @DubaiApeYC, @BoredClubCanada, @BAYCKorea, @BAYCkada in the Philippines, @BAYCJapan, @UKApeClub, @FrenchApeYC and many more countries. Most of these accounts include a link in the description for people to join their own private, dedicated Discord server for their community. Others, such as @DubaiApeYC do not have a posted link and are by invitation only.
The communities actively invite BAYC holders in the area to join and meet with others. The account description of the @BigAppleApes reads:

\textit{``We are the Apes that swing in the Concrete Jungle when we aren’t hangin in the swamp… If you're in NYC or the tri-state area, join us for our community meetup…''}

The same inviting nature is shown in the @FrenchApeYC description:

\textit{``ONLY OFFICIAL TWITTER | Group of French speaking \#BAYC \& \#MAYC owners | 150+ members | Private discord | IRL monthly events and much more …''}

The posts of these local community Twitter accounts include photos of their members at hosted events, reposts from their members’ accounts, and advertisements for in-person gatherings. The online and offline activity of these communities across the world highlight the social aspect of being a Bored Ape and reinforce the welcoming nature of the community shown in the interview data.

\subsubsection{Community Usage of IP Rights}

Yuga Labs grants a broad license with any BAYC or MAYC NFT, giving holders rights to personal and commercial use of their NFT. These usage rights have been used in various business manners. A recent development in Yuga Labs is Made by Apes, which is a Yuga Labs-supported license verified through the blockchain. The licenses are granted to authenticate that licensees own an official BAYC or MAYC NFT and that the Made By Apes logo can be used on their product. Owners already have commercial rights to their specific Ape, therefore a Made By Apes license is not necessary for business ventures. However, the license has the function of amplifying the products and services that a brand offers with the backing of the BAYC name. On Twitter, the hashtag \#MadeByApes is used to advertise Bored Ape-affiliated brands. Licensed brands range from food and beverage to video games using Apes as characters.

Some members of the BAYC also participate in creative ventures with their individual IP rights. An Ape holder on September 4th tweeted:

\textit{``I like to view my forever hodl [hold] pfps as individual characters and create bits of story around them, though few seem to care about that Well at least I can count on @MythDivision to visualise and tell their stories for me!''}

Myth Division is a non-Yuga-affiliated company dedicated to utilizing clients’ NFTs in comic books, stories, and physical collectibles. Ownership of a BAYC NFT grants the ability to own exclusive rights to one's Ape, but also to creative ventures such as Ape-affiliated businesses or artwork. It has also contributed to the development of NFT-utilizing businesses like Myth Division by allowing BAYC holders to explore novel uses for their ownership.

\subsection{Reddit}

The BAYC community on Reddit consists of 2,500 members and is located under the subreddit r/BoredApeYachtClub. This subreddit is a community-run unofficial platform, unlike Discord and the official Twitter account which are sanctioned by Yuga Labs. Posts on the subreddit largely consist of screenshots of official BAYC Twitter posts and records of BAYC and MAYC NFT sales, including information on who they were sold to and the cost in Ethereum. Not much activity was noted during the allotted research observation period, but the existence of a non-affiliated Yuga Labs social media platform suggests a cohesive community that is passionate about spreading the most recent news of the BAYC NFT project.

\section{Discussion}
\label{sec:discussion}

The narratives of BAYC collectors' experiences expand the current literature on stakeholders in the NFT ecosystem \cite{Sharma:2022}. The topics included are the first academic explorations of NFT collectors. Below, we reflect on both positive and negative experiences in the BAYC community, connecting to the hype and collapse of the NFT marketplace.

\subsection{Causation of the Initial Rise in NFT Popularity}
\label{subsec:causation-of-the-initial-rise-in-nft-popularity}

One implication of the findings, regarding the theme of mutual support of BAYC collectors, is that the online NFT community may have played a large role in the initial rise in popularity of the NFT phenomenon.
The phenomenon has sustained its extensive growth through the usage of social media platforms to create community spaces for collectors \cite{Casale-Brunet:2022}. Features of blockchain such as verification of ownership and decentralization could act as possible lures for collectors. However, the findings of this study suggest that one of the pull factors may be the welcoming nature of the online NFT community, especially for new members and NFT novices. Our participants raved about the community influencing their choice to ``ape in,'' i.e., to buy a BAYC NFT. When participants were met with the large online social presence as P4 described, it encouraged them to continue engagement in the community. This feedback-engagement cycle promoted the growth of the BAYC community.

After BAYC holders are motivated to engage with the community, they can enjoy the exclusionary benefits BAYC NFTs offer them. For example, they can express their personalities, identities, and status through BAYC NFTs. BAYC holders are also granted exclusive access to online and offline events.

\subsection{In Blockchain We Trust?}
\label{subsec:in-blockchain-we-trust?}

The NFT market essentially crashed in 2022, after the price of the digital currency Bitcoin began to crater. Many NFTs, including BAYC, lost an average of 92 percent of their value \cite{Padtberg:2022}.

The presence of encountered challenges in the findings presents a refutation of the literature-established trustworthiness of blockchain technology with its consumers. The accessibility of the Ethereum blockchain helps both legitimate and ``cash grab'' NFT projects to flourish, and the latter could make collectors lose their investment.
The creation of scams is easier than ever with the decentralized nature of the blockchain. Therefore, to NFT collectors, the nature of the blockchain comes as a double-edged sword. Protecting collectors from scams is difficult to address due to the decentralization that defines NFTs. However, there are alternative approaches to avoiding scams while keeping the foundational technology of the blockchain, e.g., community-wide communication as suggested in \cite{Sharma:2022} which was confirmed by our observational study. The supportive and social environment of the BAYC community as shown in our findings can be used to warn members of potential scams. When a collector encounters a scam, the online community can be used to the advantage of alerting a wide audience.


\subsection{Limitation and Future Work}
\label{subsec:limitations-and-future-work}
There are several limitations of our study. First, the sample size of the interview study is relatively small (N=4) and the observational study only spanned one week. Gathering interview participants was difficult possibly due to the caution BAYC collectors received about scams. We do not claim to provide generalizable or representative conclusions about the BAYC collector population but rather intend to display the particular experiences of some BAYC holders within the BAYC community. A possible avenue of further research on BAYC collectors could include large-scale quantitative surveys to obtain more generalizable insights regarding their experiences.

Second, BAYC collectors who are more involved in the NFT community, such as moderators of the BAYC Discord channel and people directly affiliated with Yuga Labs, are difficult to contact online.
DM options may be turned off by them, effectively barring the researcher from contact. A follow-up study with them could generate more insights about the governance aspect of NFT communities.

Last but not least, all participants accepting the interviews identified as male. Further works could research female BAYC NFT holders, or more generally, the demographic makeup of NFT holders. More research into the demographics of the blockchain space could lead to an explanation of the gender disparity and accessibility issues present in this domain \cite{Bannier:2019,zhou2023iterative,zhou2022toward}.

\section{Conclusion}
\label{sec:conclusion}
This study, which included interviews with four BAYC holders alongside an observational study, aimed to give a voice to the people involved in the decentralized NFT world, as the existing literature surrounding NFTs focused more on the technology or artistic aspect. Participants expressed general excitement, appreciation, and gratitude for the online NFT community, allowing them to create lasting relationships and business ties. However, this optimism periodically encountered challenges that acted as demotivators for engaging in the community, particularly through ``trolls'' and scams.
We reflect on both positive and negative experiences in NFT communities and connect them with the rises and falls of the NFT market.

\bibliographystyle{splncs04}
\bibliography{BAYC}

\appendix
\section{Interview Protocol}
\label{interview}
These questions were based on the work of Sharma et al. \cite{Sharma:2022} and other extant literature on the topic.

\subsection{Demographics}
\begin{itemize}
    \item What is your age? 
    \item What is your gender identity? 
    \item What are your preferred pronouns?
    \item What is your highest level of education?
\end{itemize}

\subsection{NFT/BAYC Engagement}
\begin{itemize}
    \item How did you first learn about NFTs? 
    \item Describe your first direct experience with NFTs. 
    \item When did you buy your first Bored Ape Yacht Club NFT?
    \item Can you remember and explain the reason for buying a Bored Ape Yacht Club NFT? 
    \item What differentiates the BAYC NFT collection from other NFT collections?
    \item How did the community aspect influence your choice of purchase?
\end{itemize}

\subsection{NFT/BAYC Community Engagement}
\begin{itemize}
    \item Are you involved in other NFT communities besides the BAYC community?
    \item What was the reason for participating in the other communities?
    \item What do you think differentiates the BAYC community from other NFT communities?
    \item How involved would you describe yourself within the Bored Ape Yacht Club community?
    \item What do you think are the benefits of participating in the BAYC community? Negatives?
    \item Have you observed common topics, values, and norms in the BAYC community?
    \item Are there any experiences you have had that demotivate you from NFT engagement or participating in online communities?
    \item Is there any other information relating to this topic that you would like to share?
\end{itemize}

\end{document}